\documentclass[%
 reprint,
superscriptaddress,
 amsmath,amssymb,
 aps,
]{revtex4-1}

\usepackage[utf8]{inputenc}
\usepackage[T1]{fontenc}
\usepackage{array}
\usepackage{textcomp}
\usepackage{graphicx}
\usepackage{dcolumn}
\usepackage{bm}
\usepackage{float}
\usepackage{xcolor}

\begin{document}

\title{Designing and manufacturing of interference notch filter with a single reflection band}

\author{Mohammadreza Salehpoor}
\affiliation{Department of Physics, University of Isfahan, Hezar Jerib, Isfahan 81746-73441, Iran}

\author{Hossein Vahid}
\affiliation{Department of Physics, University of Isfahan, Hezar Jerib, Isfahan 81746-73441, Iran}

\author{Ali Heidary~Fard}
\affiliation{Department of Physics, University of Isfahan, Hezar Jerib, Isfahan 81746-73441, Iran}

\author{Hamidreza Fallah}
\affiliation{Department of Physics, University of Isfahan, Hezar Jerib, Isfahan 81746-73441, Iran}
\affiliation{Quantum Optics Research Group, Department of Physics, University of Isfahan, Hezar Jerib, Isfahan 81746-73441, Iran}

\author{Morteza Hajimahmoodzadeh}
\affiliation{Department of Physics, University of Isfahan, Hezar Jerib, Isfahan 81746-73441, Iran}


\begin{abstract}
The main idea of this work is to design a notch filter structure with a narrow notch width and maximum reflection while reducing fabrication challenges. In addition, using anti-reflection layers in the outermost part of the designed structure, the pass-band ripples are reduced. In designed systems, various quarter-wave Coefficients were used, which have less challenge in the manufacturing part. Furthermore, the stability of the deposition conditions and the density of the layers affect their quality and consequently the result of environmental tests. Therefore, the sputtering method with RF and DC sources was used to construct the designed structure.

\end{abstract}

\maketitle

\section{Introduction}
During the last decades, notch filters (NFs) have been used in different signal processing~\cite{minguillon2017} applications, including speech and image processing, data transmission, seismology~\cite{lee2013optical}, biomedical engineering~\cite{juan2013} and other applications like Raman and fluorescence spectroscopy, laser systems, laser protective coatings~\cite{janicki2005, sprague2004}, Protection of optical equipment against high beams~\cite{verly2007} in order to control the wavelength of light and eliminate some frequencies~\cite{nikolic2018, nehorai1985, jeong2018}.
NFs are optical instruments based on the interference phenomenon, which  rejects frequencies inside a narrow band and allows other frequencies to pass without change~\cite{scherer2008, zhang2012, thelen1971, thelen1993, khazraj2017}.
 
Single-band and multi-band are of two main kinds of NFs that have been mostly used in industrial engineering. 
The main methods that have been used to produce such filters are rugate~\cite{lyngnes2014}, discrete layer (HL) designs~\cite{janicki2005, rahmlow2006}, and holographic technologies~\cite{sprague2004}. 
The rugate designs with a small difference between refractive index of layers, which have less ripple in the transmission regions~\cite{southwell1989}, are well known in the literature. The latter design requires the deposition of layers by a refractive index gradient or so-called flip-flop structures, which is very complex and difficult to produce and economically not useful. 

In the discrete layer (HL) method, materials with different refractive indices are used to manufacture NFs. H and L indicate the quarter-wave layer with the higher and lower refractive index, respectively. The optical thickness of these layers is a quarter of the central wavelength~\cite{janicki2005}. 
The challenging parts of this design are the deposition of the layers in a repeating order with high accuracy, reducing transmission band ripples, making high layer quality. In addition, low level of particle defects, reasonable stress in the layers, and less manufacturing cost per sample are among important factors.
In this method, the layers are placed on top of each other periodically that leads to ease in the design and manufacturing, but we should note that the transmission band ripples increase due to the sudden change of the refractive index between the layers with high and low refractive index. 
on the other hand, with fewer layers can be reached to a certain level of reflection in the center of the reflection band~\cite{schallenberg2010, scherer2008}.

The deposition method and selecting the appropriate materials are very effective in the quality of the filter made. The reflectance bandwidth decreases by reducing the difference between the refractive index of the two materials in the structure. 
In this study, we use appropriate materials in the HL structure to provide a functional design with high reproducibility for the NFs at a center wavelength of $532$ nm. 
Then, using anti-reflect layers, we reduce the transmission band ripples. We use the sputtering method with two RF and DC sources to manufacture the designed structure. Finally, we present the result of quality assessments conducted on the fabricated samples.

\section{Design methodology}
\subsection{Initial design}
The first step to create an optical instrument is to choose a suitable design method. An important issue that arises before selecting a design method is how to perform the final design practically. The method of quarter-wave stack structure, in the deposition stage, is more straightforward than other methods, and fewer layers are used in the structure, which make it more economical and less challenging. Furthermore, fewer layers also reduce the stress between layers and lead to better resistance to damage from environmental conditions.

In this method, the layers are placed alternately to form structures such as (nHmL)$^s$ or (nLmH)$^s$. In this notation, m and n determine the number of layers (odd integer) of each quarter-wave, and s expresses the number of pairs of layers used in the design. The reflection bandwidth or notch width (NW), as one of the characteristics of NF, depends on the difference between the refractive indices of the two materials with high and low refractive indices or the ratio between them. Therefore, the selection of materials should be such that the difference between the refractive indices of two materials optimize NW and the maximum reflection to desired values. The NW is given by~\cite{macleod2017}

\begin{equation}
    {\rm NW}=\frac{4}{\pi}\arcsin[ \frac{(\rho^{2}+2\rho\cos\gamma+1)^\frac{1}{2}}{\rho+1}],
    \label{eq1}
\end{equation}
where 
\begin{equation}
    \gamma=(\frac{s-1}{s})\pi        \quad\mathrm{and}\quad      \rho=\frac{n_{\rm H}}{n_{\rm L}}.
    \label{eq2}
\end{equation}
Equation~(\ref{eq1}) then, in a simpler form, becomes~\cite{macleod2017}
\begin{equation}
    {\rm NW}=2\Delta g=\frac{4}{\pi}\arcsin( \frac{n_{\rm H}-n_{\rm L}}{n_{\rm H}+n_{\rm L}}).
    \label{eq3}
\end{equation}

\begin{figure}[hbt!]
 \centering
\includegraphics[width=1\linewidth]{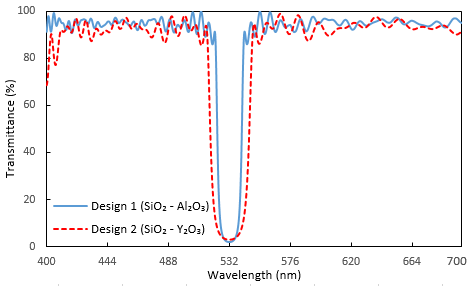}
 \caption{Transmittance diagrams of designed structures with SiO$_2$-Y$_2$O$_3$ and SiO$_2$-Al$_2$O$_3$. The information about the structures is listed in Table~\ref{tab1}.}
 \label{pic1}
\end{figure}

The maximum reflection depends on the number of repetitions of the main period, s, in the structure~\cite{macleod2017, diehl2003} and can be calculated by~\cite{macleod2017}
\begin{equation}
    R_{\rm max}=\frac{n_{\rm sub}-\rho^{2s}n_{\rm H}^2}{n_{\rm sub}+\rho^{2s}n_{\rm H}^2},
    \label{eq4}
\end{equation}
where $n_{\rm sub}$ is the the refractive index of the substrate.
In this study, we used Y$_2$O$_3$ and Al$_2$O$_3$ as the materials with high refractive index (H) and SiO$_2$ as the material with low refractive index (L).
We used BK7 with the refractive index of $1.52$ as the substrate, and we considered air with that of $1$ as the incidence medium in the design. The central wavelength of design is $\lambda=532$ nm, which is located in the center of the reflection band. The physical thickness (PT) of the layers is such that the layers have optical thicknesses equal to odd integer coefficients of one-quarter of the central wavelength. Details of the designed structures with two pairs of materials, Y$_2$O$_3$-SiO$_2$ and Al$_2$O$_3$-SiO$_2$, are given in Table~\ref{tab1}.
\begin{table}[htbp]
\centering
\caption{\bf Details of designed structures with SiO$_2$-Y$_2$O$_3$ and SiO$_2$-Al$_2$O$_3$}
\begin{tabular}{ccc}
\hline
Design number &  1 & 2 \\
\hline
$\lambda$ (nm) & $532$ & $532$ \\  
Materials & SiO$_2$-Al$_2$O$_3$ & SiO$_2$-Y$_2$O$_3$ \\
Number of layers & $44$ & $26$ \\
$R_{\rm max}$ (\%) & $98.84$  & $98.53$ \\
NW (nm) & $17$  & $26$ \\
Total PT ($\mu$m)  & $14$ & $8$ \\
\hline
\end{tabular}
  \label{tab1}
\end{table}
Although using combination of $3$ and $5$ coefficients instead of single quarter-wave layers increases the thickness of structures, but significantly reduces the transmission band ripples and the number of total layers required in the design. the transmission diagrams of both structures are given in Figure~\ref{pic1}.
From Fig.~\ref{pic1} and the information of Table~\ref{tab1}, it can be concluded that reducing the difference between the refractive indices of the two materials reduces the NW. Meanwhile, we need more layers to achieve the maximum reflection.

\subsection{Final design optimization}
We use anti-reflect layers, which are made of the same materials used in the main structure, in the outermost layers of the design to reduce ripples more. Regarding this method, the strong scattering of the equivalent refractive index of the multilayer is compensated by the scattering of the anti-reflection structure and the transmission band ripples are improved~\cite{shankar2014}. Table $2$ shows the results of the initial and optimized designs.
\begin{table*}[htbp!]
\centering
\caption{\bf Comparative performance of NF with various stack. The table presents the stack formula, maximum reflection $R_{\rm max}$, average ripple reflectivity ARR, and the notch width NW.}
\begin{tabular}{ccccc}
\hline
Design number & Stack formula & $R_{\rm max}$ (\%) & ARR (\%)  & NW (nm)  \\
\hline
 1&[(5H5L)$^4$(3H3L)$^{15}$(5H5L)$^3$]  & $98.05$ & $6.11$ & $17$   \\
 1&[(5H5L)$^4$(3H3L)$^{15}$(5H5L)$^3$]$2$HLHL& $98.84$ & $3.02$ & $17$  \\
  2&[(5H5L)$^2$(3H3L)$^9$(5H5L)$^2$] & $97.04$ & $9.57$ & $26$ \\
 2&[(5H5L)$^2$(3H3L)$^9$(5H5L)$^2$]$2$HL & $98.60$ & $3.77$ & $26$\\
\hline
\end{tabular}
  \label{tab2}
\end{table*}

\begin{figure}[hbt!]
 \centering
 \includegraphics[width=1\linewidth]{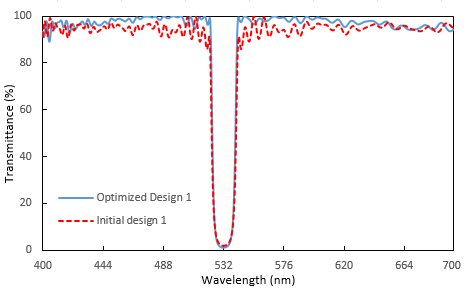}
 \caption{Initial and optimized designs of SiO$_2$-Al$_2$O$_3$. The information about structures is listed in Table~\ref{tab2}.  }
 \label{pic2}
\end{figure}

\begin{figure}[hbt!]
 \centering
 \includegraphics[width=1\linewidth]{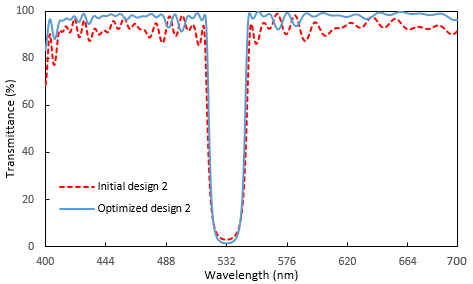}
 \caption{Initial and optimized designs of SiO$_2$-Y$_2$O$_3$. The information about structures is listed in Table~\ref{tab2}.}
 \label{pic3}
\end{figure}

In addition, the optimized transmittance diagrams for both designed structures compared to their initial designs are shown in Figures~\ref{pic2} and~\ref{pic3}. Adding anti-reflection layers increases the maximum reflection at the central wavelength.
In the next section, we discuss the feasibility of manufacturing the structure containing Al$_2$O$_3$. 
This structure has a smaller NW in regards to the smaller difference between its refractive index and SiO$_2$. Moreover, it is more resistant and cheaper than Y$_2$O$_3$.

\section{Experiment}
\subsection{Manufacturing method}
 The layers were deposited employing Reactive magnetron sputtering method, which is a suitable method for dielectric materials deposition, and it can produce dense layers with good resistance to environmental stresses. 
Increasing the deposition rate by increasing the applied power, reducing sputtering of the substrate and vacuum chamber, which prevents contamination, are the advantages of this method.
Reducing the substrate's heating during deposition enables the sputtering process to be used for different substrates. Also, this method needs less gas pressure during the deposition and is economically affordable due to its material consumption compared to other methods.

\subsection{SiO$_2$ and Al$_2$O$_3$ single layers }\label{sec_dep}
\begin{figure}[hbt!]
 \centering
 \includegraphics[width=1\linewidth]{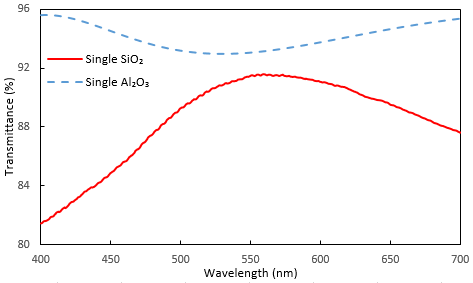}
 \caption{The spectral transmission of the SiO$_2$ and Al$_2$O$_3$ single layers.}
 \label{pic4}
\end{figure}

In the design software, we design a single layer of SiO$_2$ and a single layer of Al$_2$O$_3$ at the center wavelength of 532 nm. 
In this step, we use SF6 and BK7 as substrates for SiO$_2$ and Al$_2$O$_3$ single layer, respectively. 
The deposition was done using DC source for SiO$_2$ and RF source for Al$_2$O$_3$ with silicon and aluminum targets with 99.99\% purity and after complete cleaning of the substrate. Optimal conditions for deposition of SiO$_2$ and Al$_2$O$_3$ were obtained in 11 sccm and 3 sccm oxygen flow, respectively. 
Deposition rates of SiO$_2$ and Al$_2$O$_3$ are 6 \AA/s and $1.5$~\AA/s respectively. The deposition was performed for both single layers at 100\textdegree C, chamber pressure of 3 mTorr during deposition, the argon gas flow of 25 sccm, and the start pressure of $8\times10^{-6}$ Torr. Figure~\ref{pic4} shows the spectral transmission of the fabricated single layers.
\subsection{(3H3L)$^2$ stack}\label{sec3}
\begin{figure}[hbt!]
 \centering
 \includegraphics[width=0.9\linewidth]{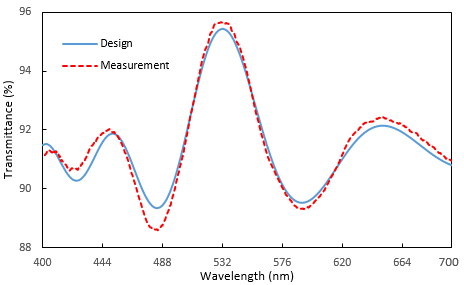}
 \caption{Measured and designed transmittance spectra for (3H3L)$^2$ stack.}
 \label{pic5}
\end{figure}

Optimal conditions for deposition of stacks are the same as ones used for single layers. However, before deposition the final design structure and in order to achieve better results, it is necessary to deposit a simpler structure with fewer layers.
Figure~\ref{pic5} shows the transmittance spectra of the deposited and the designed structures are in good agreement. It indicates that the obtained deposition conditions in the previous sections for single layers are approximately the optimal ones.
\subsection{NF deposition}
According to design 1, we use silica and alumina materials as H and L refractive indices, respectively. Regarding the optimized structure, see Figure~\ref{pic3}, the final design at the central wavelength of $532$ nm is given by the design formula number 1 (see Table~\ref{tab1}).

In Sec.~\ref{sec3}, by performing several experiments,  design and fabrication of single layers of materials in the structure, optimizing deposition conditions, and finally designing and deposition of 4--layer stacks were done. 

The deposition conditions are similar to explained ones in Sec.~\ref{sec_dep}.
The control of deposition rate and the thickness of the layers is crucial.

\begin{figure}[hbt!]
 \centering
 \includegraphics[width=0.9\linewidth]{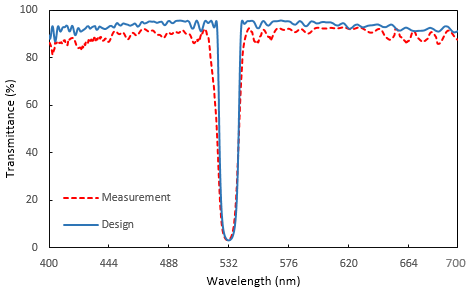}
 \caption{Measured and designed transmittance spectra for NF (SiO$_2$-Al$_2$O$_3$).}
 \label{pic6}
\end{figure}

In Sec.~\ref{sec3}, the optimization was performed for samples, which are located on the rotating segment, and the necessary corrections were made. Therefore, we expect the samples to show good compliance with the designed structure. 
In the transmittance spectra, second side of the sample has a reflection, so first in the design software, we applied the effect of the reflection of the second side of the samples, and then we compare design and measured transmittance spectra. Figure~\ref{pic6} shows the measured and designed spectra of the central sample.

As shown in Figure~\ref{pic6}, the final deposited structure and the final designed structure at the location of the central wavelength and the reflectance bandwidth are well matched. By extracting the data obtained from the transmission spectrum, the center of the reflection band is located at $530$ nm, which is a perfect match with the design. We can see the ripples only in some areas of the transmission band.  In the transmission band, there is a mismatch at some point between the transmittance spectrum of the deposited structure and the designed structure. 
Generally, the mismatches between the transmittance spectra of designed and deposited structures in the coating technology are due to the mismatches between the thickness of the same layers, the refractive indices, and the packing density in the deposited and the designed structures. Absorption in some coating areas, contamination, and error of spectrometer and operator are other factors.
In Figure~\ref{pic6}, although the spectra of the designed and the deposited structures are very well matched in many wavelengths, the control and stability of the deposition process must be very high to achieve spectra closer to the designed structure. 
Some of these mismatches are inevitable because the final designed structure of NF consists of 52 layers with different thicknesses.
In the manufactured NF, the NW was $17$ nm, the maximum reflectance at central wavelength was more than $95$ \%, and the average transmittance was approximately $90$ \%.
\section{Conclusions}
 HL method has fewer challenges in designing and manufacturing NFs. 
In this method, the differences between the refractive indices of the materials used in structures are the determining factor in the NW and the maximum reflection. In addition, combined coefficients of 3 and 5 in the design of the NFs are beneficial because these coefficients significantly reduce the transmittance band ripples compared to the single quarter-wave structures.
The ripples in the transmittance spectra are essential challenges in the design and fabrication of NFs. In this paper, by antireflection layers and a half-wave layer used as the interface layer in the outermost layers of the final structure, the transmission band ripples were reduced appropriately.
Magnetron sputtering method is a very new and efficient method with suitable economic cost. This method produces high-quality layers. Success in environmental tests depends directly on the manufacturing method. In this study, samples made by the magnetron sputtering method were subjected to environmental tests, which yielded positive results. 
It should be noted that the stability of the coating rate and precise control of the inlet gases to the chamber during the coating process has a significant impact on the formation of coatings with appropriate stoichiometry. In addition, increasing the substrate temperature causes better adhesion of the first layer to the surface of the substrate. 
\bibliographystyle{achemso}
\bibliography{references}
\end{document}